\documentstyle[12pt,aasms4,epsfig,psfig]{article}

\begin{document}

\title{Gamma-loud quasars: a view with {\it Beppo}SAX}

\author{Tavecchio F.$^1$, Maraschi L.$^1$, Ghisellini G.$^2$, Celotti
A.$^3$, Chiappetti L.$^4$, Comastri A.$^5$, Fossati G.$^6$, Grandi
P.$^7$, Pian E.$^8$, Tagliaferri G.$^2$, Treves A.$^9$, Raiteri
C.M.$^{10}$, Sambruna R.$^{11}$, Villata M.$^{10}$}

\altaffiltext{1}{Osservatorio Astronomico di Brera, via Brera 28, 20121 Milano,
Italy}

\altaffiltext{2}{Osservatorio Astronomico di Brera, via Bianchi 46, 23807 Merate, Italy}

\altaffiltext{3}{SISSA/ISAS, via Beirut 2-4, 34014 Trieste, Italy}

\altaffiltext{4}{IFC/CNR, via Bassini 15 , 20133, Milano,  Italy} 

\altaffiltext{5}{Osservatorio Astronomico di Bologna, via Ranzani 1, 40127, Bologna, Italy}

\altaffiltext{6}{CASS, University of California, San Diego, CA 92093-0424}

\altaffiltext{7}{IAS, IAS/CNR, via Fosso del Cavaliere, 00133 Roma, Italy}

\altaffiltext{8}{ITeSRE/CNR, via Gobetti 101, 40129 Bologna, Italy}

\altaffiltext{9}{Universita' dell'Insubria, via Lucini 3, 22100, Como, Italy}

\altaffiltext{10}{Osservatorio Astronomico di Torino, Strada Osservatorio 20, 10025, Pino
Torinese (TO), Italy}

\altaffiltext{11}{Pennsylvania State University, 525 Davey Lab, State College, PA 16802}

\setcounter{footnote}{0}




\begin{abstract}
We present BeppoSAX observations of three $\gamma $-ray emitting quasars,
namely 0836+710, 1510-089 and 2230+114. The three objects have been detected up
to $\sim $100 keV showing extremely flat power-law spectra above 2 keV
(energy index $\alpha _{2-10}=0.3-0.5$). The soft X-ray spectrum of 
0836+710 implies
either an absorption column density higher than the galactic one or an
intrinsically very hard slope ($\alpha _{0.1-1}=-0.2$) below 1 keV.
1510-089 shows a soft excess, with the low energy spectrum steeper
($\alpha _{0.1-1}=1.6$) than the high energy power-law. The results are
discussed in the framework of current Inverse Compton models for the high
energy emission of Flat Spectrum Radio Quasars and are used to estimate
the physical quantities in the jet emitting region and to shed light on
the energy transport mechanisms in jets. Finally we discuss the estimates
of the jet luminosity in the context of the Blandford \& Znajek mechanism
for jet production.
\end{abstract}

\keywords{quasars: general --- quasars: individual (S5 0836+710, PKS 1510-089,
HB89 2230+114) --- X-rays: spectra --- radiation mechanisms: non-thermal}

\section{Introduction}

Since the EGRET detection of about 60 blazars (BL Lac objects and Flat
Spectrum radio Quasars, FSRQ) in $\gamma$-rays (Mukherjee et al. 1997),
the study of this class of objects has received renewed interest. A large
fraction of the total power is in fact emitted in the $\gamma$-ray band,
which is therefore crucial to test different radiation models.

The bright $\gamma$-ray emission requires relativistic boosting
(e.g. Dondi \& Ghisellini 1995), confirming that the emission of blazars
originates in relativistic jets. The Spectral Energy Distribution (SED)
of all observed blazars from radio to $\gamma$-rays shows two broad
peaks, believed to be produced by relativistic electrons in the jet via
the synchrotron and Inverse Compton (IC) processes respectively.  It has
been suggested that the location of the synchrotron and IC peaks and the
``$\gamma$-ray dominance'' (i.e the luminosity ratio of the second to the
first peak) are related to the power of the source so that all blazars
lie along a spectral sequence ( Fossati et al. 1998; Ghisellini et al
1998).  FSRQs have synchrotron and IC peaks located at lower energies and
are more $\gamma$-ray dominated than the less powerful BL Lac objects.

It is widely believed that the IC emission from FSRQ is dominated by the
scattering between high energy electrons and soft photons external to the
jet (EC). The latter ones may be produced by the accretion disk itself
and be scattered/reprocessed in the Broad Line Region (BLR) (Dermer \&
Schlickeiser 1993; Sikora, Begelman \& Rees 1994). However other sources
of seed photons (e.g. the synchrotron photons themselves, SSC) could
contribute in the soft-medium X-ray band.  Contributions due to other
components, like direct synchrotron emission or the high energy tail of
the blue bump, are also possible in the soft X-ray band.  In the simplest
case of a single EC component, the flat X-ray spectrum of FSRQ represents
the low energy side of the IC peak and therefore is due to {\it low
energy electrons} scattering the externally produced photons. This
spectral band gives then information on a part of the electron spectrum
which, due to selfabsorption, is not accessible in the low frequency
synchrotron component.

The unprecedentedly wide band of {\it Beppo}SAX (from 0.1 up to 100 keV)
is optimal to study the connection between the X-ray and the $\gamma$-ray
continuum and constrain and disentangle different emission components.
For these reasons we started a program to observe the brightest
$\gamma$-ray blazars. Here we report results for three of them, namely
0836+710, 1510-089 and 2230+114. After a summary of previous observations
for each source (section 2), we present the analysis of the {\it
Beppo}SAX data (section 3). The implications are discussed in Sect. 4 in
the framework of the EC model for FSRQ. Conclusions are given in Sect. 5.

\section{The observed objects}

All sources have been repeatedly observed in $\gamma $-rays by EGRET
showing a steep spectrum. Results are summarized in Table 1 together with
previous X-ray measurements from ROSAT and ASCA. In the following we briefly 
describe the sources' characteristics most relevant for the present work.\\

\noindent
{\bf 0836+710}: This is a distant bright FSRQ.  VLBI observations
show a compact core and ejections of components with superluminal motion,
possibly connected to $\gamma $-ray flares (Otterbein et al. 1998).

In the X-ray band 0836+710 has been observed by ROSAT and ASCA (Cappi et
al. 1997). In both cases it showed a flat spectrum that, together with
the steep $\gamma $-ray emission seems to indicate (as data are not
simultaneous) that the IC peak lies in the soft $\gamma $-ray band. A
deficit of soft photons ($E<1 $ keV) was interpreted as evidence of absorption higher
than the galactic one.  Moreover from a comparison of ROSAT and ASCA
data (Cappi et al. 1997) this absorption appeared to change by $\Delta
N_H \sim 8\cdot 10^{20}$ cm$^{-2}$ on a timescale of less than 2.6 years
(0.8 yr in the quasar frame).

Recently Malizia et al. (2000) reported the detection of 0836+710 by
BATSE in the energy range 20-100 keV with a flat spectrum ($\alpha
$=0.1-0.3). 

The optical spectrum shows broad emission lines superimposed
to a broad blue bump continuum, well fitted by a black body at a
temperature $T\simeq 2.5\times 10^4$ K and a luminosity $L\simeq 10^{47}$
erg s$^{-1}$ (Lawrence et al. 1996).\\

\noindent
{\bf 1510-089}: It is a nearby highly polarized quasar (HPQ),
 which presents strong
similarities with 3C273. In particular, it shows a pronounced UV bump
(Pian \& Treves 1993). 

1510-089 has been extensively observed in X-rays by EXOSAT (Singh, Rao \&
 Vaia 1990, Sambruna et al. 1994), GINGA (Lawson \& Turner 1997), ROSAT
 (Siebert et al. 1995) and ASCA (Singh, Shrader \& George 1997).  The
 X-ray spectrum is very flat in the 2-10 keV band, while in the ROSAT
 band it is steeper, suggesting the possible presence of a
 spectral break around 1-2 keV.  The EXOSAT observation (Singh et
 al. 1990) showed the presence of a relatively strong iron line
 (EW$\simeq$800 eV), not detected in more recent observations with GINGA
 and ASCA.\\

\noindent
{\bf 2230+114}: This quasar has been observed several times
with GINGA (Lawson \& Turner 1997), ROSAT (Brinkmann et al. 1994) and ASCA
(Kubo et al.  1998).  It shows a flat spectrum which
extends smoothly into the soft $\gamma $-ray band, as indicated by the OSSE
data (McNaron-Brown et al. 1995).

Radio observations reveal a rather complicated structure (see Altschuler et
al. 1995 and references therein) and
superluminal components. Kellermann et al. (1962) classify 2230+114 as a
classical Gigahertz Peaked Source (e.g. O'Dea 1998).

Falomo, Scarpa \& Bersanelli (1994) report the optical spectrum, which shows bright
emission lines superimposed to a blue continuum, typical of a quasar
blue bump.\\

\section{{\it Beppo}SAX observations and data analysis}

The scientific payload of the Italian-Dutch X-ray satellite
$Beppo$SAX\footnote{\small http://www.sdc.asi.it}
(see Boella et al. 1997) consists of four coaligned Narrow Field
Instruments (NFIs) and two Wide Field Cameras.  Two of the NFIs use
concentrators to focalize X-rays: the Low Energy Concentrator
Spectrometer (LECS) has a detector sensitive to soft-medium X-ray
photons (0.1-10 keV), while the Medium Concentrator Spectrometer (MECS)
reveals photons in the energy range 1.3-10 keV. The Phoswich Detector
System (PDS), sensitive from 12 up to 200 keV, consisting of four
identical units, uses rocking collimators so as to monitor source and
background simultaneously with interchangeable units. We will not 
consider here the fourth NFI, a High Pressure Gas Scintillation
Proportional Counter (HPGSPC).

The {\it Beppo}SAX journal of observations is reported in Table 2, with
exposure times and observed count rates. The observation of 2230+114 
consisted of 5 short pointings, separated
by 1-2 days, while the others were continuous apart from gaps due to satellite
constraints.
None of the sources showed significant flux variations within the observations.
We therefore obtained a cumulative spectrum for each source.

We analyzed the {\it Beppo}SAX spectral data using the standard software
packages XSELECT (v1.4) and XSPEC (v10.0) and the September 97 version of
the calibration files released by the $Beppo$SAX Scientific Data Center
(SDC).  From the event files we extracted the LECS and MECS spectra in
circular regions centered around the source with radii of 8$^{\prime }$ and 
4$^{\prime }$ respectively (see the SAX Analysis Cookbook\footnote{\small
ftp://www.sdc.asi.it/pub/sax/doc/software\_docs/saxabc\_v1.2.ps.gz}).
The PDS spectra extracted with the standard pipeline with the rise-time
correction were directly provided by the $Beppo$SAX SDC. We used PDS
data rebinned with S/N$>3$.

For the spectral analysis we considered the LECS data in the restricted
energy range 0.1-4 keV, because of unsolved calibration problems at
higher energies. Background spectra extracted from blank field
observations at the same position as the source were used.  We fitted
rebinned LECS, MECS and PDS spectra jointly, allowing for two variable
different normalization factors to take into account uncertainties in the
intercalibration of different instruments (see SAX Cookbook).

In the following subsections we report the analysis procedure for each
 object. The results of the spectral fits are summarized in Table 3.

\subsection{0836+710}

The total LECS+MECS+PDS spectrum of 0836+710 is well described by a
single absorbed power-law model, with a column density higher than the
galactic value at the 99\% confidence level (see Figs \ref{0836_plnhfree}
and \ref{0836_multi}). Assuming a model with fixed galactic absorption
plus a free absorption at the redshift of the source (the ZWABS
model of XSPEC) we found that the required intrinsic column density is
$N_H=6.6 \,(2.5-12.6)\times 10^{21}$ cm$^{-2}$ to the quasar rest frame.

An alternative possibility is to model the spectrum with a broken power
law. This gives an extremely flat low energy slope, with an acceptance
probability equivalent to the single power law plus free absorption
model.  These results agree well with the findings of Cappi et al. (1997)
from ASCA data even at different flux levels (see the discussion in
Sect. 3.1.1).

The residuals of the fits (see Fig. [\ref{0836_plnhfree}]) show an excess
around 2 keV in the LECS data. This is just the expected energy of the
redshifted fluorescence Fe line at 6.4 keV. Fixing the continuum
parameters to the value given by the fit of LECS, MECS and PDS data for a
single power-law we tried to model the excess with an unresolved gaussian line
with the energy as a free parameter. The fit converges to the right
energy ($E=2.0 \pm 0.1$ keV, EW$\simeq 110$ eV), but the improvement in
the $\chi^2$ is only marginal (the $F$--test gives a probability $P\sim
90\%$). No excess is present in the MECS spectra at the same energy,
which however is near the low energy end of the MECS sensitivity band,
fixed at 1.8 keV (as indicated in the SAX Cookbook).  We checked whether
the feature could be due to extraction problems or contamination: indeed when
extracting the LECS spectrum in a smaller region of 4' the residuals are
less evident.  We conclude that the emission feature is probably not
real.

\subsubsection{Comparison with ROSAT observations}

In Fig. (\ref{0836_multi}) we report the $N_H$--photon index confidence
contours of ROSAT, ASCA and $Beppo$SAX observations, while in
Fig. (\ref{0836_nh}) we show the absorption column obtained from
different observations as a function of the 0.1-2 keV flux. As noted
above, the SAX and ASCA confidence contours in Fig. (\ref{0836_multi})
are clearly consistent, implying an $N_H$ higher than galactic and a very
flat continuum ($\alpha\sim 0.3$) although the flux during the SAX
observation was larger than that measured by ASCA by a factor of 2. On
the contrary, the ROSAT data indicate a steeper spectrum ($\alpha \sim
0.5$) and do not require extra-absorption. In fact comparing the ROSAT
and ASCA results Cappi et al. (1997) concluded that the intrinsic
absorption had varied between the two observations.  Given the
consistency of the ASCA and $Beppo$SAX results at different flux levels
we believe that the discrepancies with the ROSAT results may be more
plausibly explained by calibration problems of ROSAT (see e.g. Iwasawa, Fabian
\& Nandra 1999).

\subsection{1510-089}

A fit over the whole range (LECS+MECS+PDS) with a single absorbed
 power-law model, although statistically acceptable, produces evident
 excess residuals at low energies and in the PDS band (see
 Fig. [\ref{1510}], upper panel). Using a broken power law model, with
 the low energy photon index steeper than the high energy one, we obtain
 a better fit and the low energy residuals disappear (Fig.[\ref{1510}],
 lower panel). The significant improvement is confirmed by the $F$-test
 (probability $>99.9 \%$). Moreover this model is consistent with
 both ROSAT and ASCA observations described in Sect.1, which show a large
 difference in the hard and soft spectral indices. The data are also
 consistent with a fit with a power law+black body model, the latter
 having a temperature $kT\simeq 0.2$ keV ($\chi^2=39.2$, 56 d.o.f.).  It
 is interesting to note here that a similar soft excess has been detected
 in 3C273 by EXOSAT, ROSAT and $Beppo$SAX (Turner et al. 1990, Leach,
 McHardy \& Papadakis 1995, Laor et al. 1994 and Grandi et al. 1997, but
 see Haardt et al. 1998).

The fit of the three datasets (LECS+MECS+PDS) with a broken power law and
free intercalibration factors still gives an unacceptable value of the
PDS/MECS normalization: the best fit value is 1.84, with a 90\%
level confidence range of 1.3-2.5, while the usual value for the PDS/MECS 
relative normalization is 0.85 (90\% level confidence range of 0.77--0.93
, see SAX Cookbook).  If we fix the normalization to 0.85 the PDS data
fall well above the MECS extrapolation (see Fig. [\ref{1510}]).  The
reason for this discrepancy is possibly the contamination by another
source which lies in the PDS field of view (we remind that the PDS FOV is
quite large, $\sim 1$ deg). The ROSAT image centered on 1510-089 shows a
very crowded field and in a radius of 40$^{\prime }$ we found 6-7 sources 
with a
ROSAT flux of 1/10 of the flux of 1510-089. In fact it is possible that
1510-089 lies in a poor cluster (see e.g. Yee \& Ellingson 1993). The
integrated flux from these sources in the ROSAT band is $\sim$2/3 of the
flux from our target. Since PKS 1510-089 is intrinsically quite hard
($\alpha\simeq 0.4 $ in the MECS range) and taking into account that the
PDS effective area off axis is on average 1/2 of the on-axis area,
this source complex cannot contribute significantly in the hard band if
it has an average spectral slope $\alpha\simeq 1$. However it is
possible that one of these or other sources may be strongly obscured in
the soft-medium X-ray band, giving rise to a significant contribution in
the PDS band despite its weakness in the ROSAT field. Therefore we cannot
exclude that the PDS excess is due to contamination from another object,
although we are not able to identify it.

Using EXOSAT data Singh et al. (1990) found the presence of a relatively
strong fluorescence Fe K$\alpha $ line (EW=800$\pm$400 eV), confirmed by
Sambruna et al. (1994). The residuals in Fig.(\ref{1510}) do not show
evidence of this line: fixing the parameters of the continuum, the energy
of the line and using a gaussian profile we found an upper limit to the
intrinsic EW of $80$ eV, well below the value given by the EXOSAT data,
which refer to a similar intensity state. Interestingly this upper limit
is consistent with that found in 3C273 by Grandi et al. 1997 (but see
Haardt et al 1998).

\subsection{2230+114}

For 2230+114 we obtain a good fit of the LECS+MECS+PDS spectra with a
flat absorbed power-law model. There is no evidence for spectral breaks
or steepening and the absorption column is consistent (within 1$\sigma $)
with the galactic value (see the Fig. [\ref{2230_cont}]).  The residuals
do not show evidence of spectral features.

\section{Interpretation}

Using SAX X-ray , quasi-simultaneous optical and historical data taken
from the literature we have assembled the SED shown in Figs
[\ref{0836_sed}], [\ref{1510_sed}] and [\ref{2230_sed}] (see the figure
captions for the references to the data). In each case the $\gamma $-ray
points are averages over the available observations as given in the 3rd
EGRET catalog (Hartman et al. 1999). For the case of 2230+114 the
available OSSE and COMPTEL data are not shown since they partly overlap
with the more accurate and simultaneous data from the PDS. The
simultaneous optical data were taken at the Torino Observatory with the
1.05 m REOSC telescope. Magnitude calibration was performed according to
the photometric sequences in Villata et al. (1997) and Raiteri et
al. (1998).

The X-ray data clearly trace the low energy branch of the high energy
component, whose peak frequency falls above 1 MeV, as estimated using the
EGRET data (although not simultaneous). Note that the position of the
synchrotron peak is very uncertain, because of the poor coverage in the
IR band.

Currently the broad band properties of quasar-like blazars are understood
in the framework of the EC models.  As suggested by the strong thermal
features present in the optical-UV spectra (e.g. the blue-bump, bright
emission lines) the quasar environment is rich in soft photons, produced
by the accretion disk and/or by the BLR. In these conditions, in the
frame of the emitting source, the energy density of the external soft
radiation $U_{ext}^{\prime}$ can be much higher than the energy density
of the synchrotron radiation, and therefore the EC emission can dominate
over the SSC emission. \\

The simplest scenario to account for the emission assumes a spherical
homogeneous source filled by relativistic electrons and is fully
specified by 9 parameters: the size of the emitting region, the Doppler
factor, the magnetic field, the energy density and the frequency $\nu
_{ext}$ of the external radiation field and the parameters of the
electron distribution.  As indicated by the spectral shape of the IC
component, flat in X-rays and steep in $\gamma$-rays, the latter can be
approximated by a broken power-law, specified by two spectral indices,
the break energy and the normalization.

The observed SED in principle can yield 6 quantities, namely the
synchrotron peak frequency and luminosity, the IC
peak frequency and luminosity and the spectral
indices of the IC component (directly connected to the
indices of the electron distribution). It is worth
anticipating that for these blazars  the observed peak
corresponds to the synchrotron selfabsorption frequency of the synchrotron 
component which is higher than the peak frequency corresponding 
to the unabsorbed spectrum.

In addition to these 6 quantities the typical variability timescale can
give an indication of the size of the source. Also in specific sources
(e.g. 0836+710) it is possible to have information on the luminosity and
the typical frequency of the external radiation field. If the size of the
region where this radiation is diluted can be estimated (e.g. with the
$R_{BLR}-L_{BLR}$ relation for radio-quiet quasars, e.g. Kaspi et
al. 2000), the total number of observational constraints (6+3) is equal
to the number of parameters of the model which can therefore {\it in
principle} be {\it strongly constrained}.

\subsection{Spectral fits}

In order to reproduce the observed SEDs we adopted the model described in
Ghisellini et al. (1998): relativistic electrons with a power-law
distribution above a minimum energy are continuously injected in a
spherical emitting region with radius R, with a magnetic field intensity
B and moving with bulk Lorentz factor $\Gamma $ (in the following we
assume that the observing angle is $\theta\simeq 1/\Gamma $ and therefore
$\delta \sim \Gamma$). The injected electron distribution is
characterized by the spectral index $n_{\rm inj}$, the minimum Lorentz
factor of particles, $\gamma _{\rm min,inj}$, and the injected luminosity
$L_{\rm inj}$.  Electrons cool rapidly through synchrotron and IC losses
reaching an equilibrium distribution, which is essentially a broken
power-law with indices $n_2=n_{\rm inj}+1$ above the break at $\gamma
_{\rm min,inj}$ and $n_1=2$ below the break (in the absence of escape,
pair production and Klein-Nishina effects) down to a minimum Lorentz
factor $\gamma_{\rm min}$.

The spectrum of the soft external photons is described by a black body
with $\nu _{ext}\simeq 1-2 \times 10^{15} $ Hz and luminosity $L_{\rm
BLR}$ diluted in a spherical region with radius $R_{BLR}$. $L_{\rm BLR}$
can be related to the luminosity observed from the disk by $L_{\rm BLR} =
f L_{\rm disk}$ where $f$ represents the fraction of disk luminosity
reprocessed in the BLR.  We do not consider radiation coming directly
from the accretion disk, which at distances involved here ($\sim 10^{17}$
cm) and for $\Gamma=5-10$ is strongly redshifted.

The parameters required to reproduce the observed SEDs
(Figs. [\ref{0836_sed}], [\ref{1510_sed}] and [\ref{2230_sed}]) are given
in Table 4. It is interesting to note that the luminosity and size of the
BLR are consistent with the observed disk luminosity for
$f=10^{-1}-10^{-2}$ and a size $R_{\rm BLR}$ of a fraction of a pc. The
size of the emitting region and the Lorentz factor have been chosen so as
to allow variability with a day timescale as often observed in $\gamma
$-rays.

In the SED model shown in Figs. (\ref{0836_sed}), (\ref{1510_sed}) and
(\ref{2230_sed}) the peak of the synchrotron component is determined by
self-absorption.  Consequently at low frequencies the model has the
standard self-absorbed spectrum with slope 5/2. Additional emission
components from regions further out in the jet are necessary to account
for the spectra at lower frequencies, as indeed expected if the flat
radio spectra of blazars are due to the superposition of different
selfabsorbed components from different locations in the jet
(e.g. Begelman, Blandford \& Rees 1984). The SEDs calculated here refer to
the innermost emitting region.

An interesting feature of the EC model is the presence of a spectral
break in the soft X-ray continuum, which reflects very sensitively the
minimum energy of the scattering electrons: $\nu _{ICB}\simeq \nu_{\rm
ext} \Gamma^2 \gamma ^2_{\rm min}$ (e.g. Sikora et al. 1993). For
$\Gamma\simeq 10$ and $\gamma _{\rm min}\simeq 1$, $\nu _{ICB}\simeq
10^{17} \nu _{\rm ext,15}$ Hz.  For frequencies below $\nu _{ICB}$ only
seed photons with $\nu < \nu _{\rm ext}$ are available so that the
Compton spectrum will be depleted (flatter) with respect to the spectrum
above $\nu _{ICB}$, where the bulk of the seed photons is scattered
(Sikora et al. 1997, Ghisellini 1996).  The lack of soft X-ray photons
observed in 0836+710, previously interpreted as due to the presence of
intrinsic absorption (e.g. Cappi et al. 1997), could be due to a curved
spectrum produced by this effect (see Fig. [\ref{0836_sed}]). Note that
the position of this spectral break strongly constrains $\gamma_{\rm
min}$ to be $\leq$ a few. This is relevant for the study of the global
energy and matter content in jets (e.g. Celotti \& Fabian 1993, Celotti
et al. 1997, Sikora \& Madejski 2000).

In general the IC spectrum will be the sum of the SSC and EC
emission. For 1510-089 the SSC peak lies at energies between the UV and
the soft X-ray band and its presence could account for the observed soft
X-ray excess in the soft X-ray band in this source. Alternatively this
excess could be the hard tail of the observed strong UV bump as proposed
for 3C273. While both the above explanations appear plausible, it is also
possible that the excess could be produced by IC scattering of external
photons by a population of cold electrons as discussed by Begelman et
al. (1987) (se also Sikora et al. [1997]). For the other two sources the
SSC contribution lies well below the EC spectrum.

The adopted model predicts a slope close to 0.5 on the low energy portion
of the IC bump, due to radiatively cooled electrons below the minimum
injection energy. In particular for 0836+710 the data require a flatter
power law. This might be attributed to the fact that electrons escape
before cooling and/or that a reacceleration process energises the cooled
particles.

\subsection{Energy transport in jets}

The estimate of the physical parameters in the emitting region of
$\gamma $-loud quasars allows us to calculate the relevant energy densities
and corresponding flux along the jet. This was done initially by
Celotti \& Fabian (1993) and for a larger sample by Celotti et al. (1997).
The total transported energy flux can be expressed as:

\begin{equation}
L_k=\pi R^2 \Gamma ^2 ( U^{\prime }_B+U^{\prime }_e+U^{\prime }_p)\beta c
\label{enflux}
\end{equation}

\noindent
where $U^{\prime }_e$, $U^{\prime }_p$ and $U^{\prime }_B$ are the rest
frame energy densities of relativistic electrons, protons and magnetic
field, respectively. $U^{\prime }_e$ can be expressed as $U^{\prime
}_e=n_e<\gamma >m_ec^2$, where $n_e$ is the numerical electron density
and $<\gamma > $ is the average Lorentz factor. Radiation is excluded
from $L_k$ since the jet is optically thin.

In a jet composed by an electron/positron plasma $U^{\prime }_p=0$, while
for a proton/electron plasma with no cold electrons $n_p=n_e$ and if
protons are assumed to be cold, $U^{\prime }_p/U^{\prime }_e=m_p/(m_e
<\gamma>) $. In Tab. 5 we report the energy densities and the energy
fluxes estimated from the spectral model.

We report in line 3 of Tab. 5 the total emitted power, i.e. the observed
power integrated over the whole solid angle (e.g. Sikora et al. 1997).

It appears from Table 5 that the magnetic energy density dominates over the
energy density of relativistic electrons but not by a large factor,
i.e. it is reasonably close to equipartition.

The radiated power is much larger than that carried by relativistic
electrons and by the Poynting flux associated with the field in the
emission region. It seems then necessary to postulate that the proton
component dominates the energy transport or that the jet is structured so
that a large Poynting flux is carried outside the emitting region.  If
the number density of protons was the same as that of electrons, the
proton kinetic power would reach values of the order of $10^{47}-10^{48}$
erg/s. We recall that this estimate is based on the fact that from the
observed X-ray emission we can actually constrain the number of low
energy electrons.  The proton contribution would however be overestimated
if the jet was partly composed of {\it relativistic }pair plasma. A large
component of {\it cold} pairs is excluded by our data (except possibly
for 1510-089), since it would give rise to an excess around 1 keV which
is not observed (e.g. Sikora \& Madejski 2000).

The powers transported by these jets are quite large even when the proton
 component is not included. It is interesting to compare them with those
 released by the accretion process which can be estimated directly from
 the strength of the blue bump which is measured in all three objects.
 The derived accretion disk luminosities are listed in the last column of
 Table 5. For 0836+710 we used the luminosity of the blackbody spectrum
 fit to the data by Lawrence et al. (1996), while for 1510-089 and
 2230+114 we estimated $L_{\rm disk}$ from the UV (Pian \& Treves 1993)
 and the optical spectrum (Falomo et al. 1994), respectively. It is
 remarkable that in each case the disk luminosity is of the same order of
 that  radiated
 by the associated jet. Note that in the hypothesis $n_p=n_e$ the power
 transported in the jets would exceed the luminosity emitted by the accretion
 disks by a large factor.

Let us briefly discuss the above numbers in the context of the Blandford
\& Znajek (1977) scenario of powering the jet from the black hole
spin. In this model the jet power depends on the black hole mass and
angular momentum and on the intensity of the magnetic field threading the
horizon.  Recently the relation between the magnetic field at the
innermost stable orbit of the disk and that at the black hole horizon has
been discussed in depth by (Ghosh \& Abramowitz (1997) (GA) Livio,
Ogilvie \& Pringle (1999) (LOP). Using the standard Shakura \& Sunyaev
(1973) model for the accretion disk GA arrived at the following
expression for the extractable power valid for high accretion rates, when
the disk is dominated by radiation pressure:
\begin{equation}
L_{\rm BZ}=2\times10^{44}M_8a^2 {\rm \,\,\,\,\, erg\, s}^{-1}
\label{garad}
\end{equation}
\noindent
where $M_8$ is the black hole mass (in units of $10^8$ $M_{\odot }$) and
$a$ is the spin parameter ($a=1$ for a maximally rotating black hole).

Comparing this expression with the values found above, we conclude that
even for maximal values of the accretion rate, spin parameter and mass
($10^9 M_{\odot }$), the estimated powers are insufficient for at least two
objects. If protons are included the discrepancy becomes extremely large.
We are therefore led to suggest that the analysis of GA and LOP
probably represents a lower limit to the jet powers produced.  In fact
Krolik (1999) points out the difficulties of a realistic approach to 
the determination of the magnetic field near the black hole horizon,
while Meyer (1999) considers the possibility of a magnetic field
amplification due to the frame dragging in the ergosphere of a Kerr hole,
leading to jet power estimates larger by a factor 100 than those given by
GA. Allowing for this uncertainty we introduce a parameter $\xi$ on the
right hand side of eq. 2.

Using now the observed disk luminosity we can write:

\begin{equation}
L_{\rm disk} \simeq 1.3\times10^{46}\dot{m}M_8 {\rm \,\,\,\,\, erg\, s}^{-1}
\end{equation}
\noindent
(where $\dot{m}$ is the accretion rate in Eddington units,
$\dot{m}=L/L_{\rm Edd}$) which, combined with eq.(\ref{garad}) yields
$\xi M_8 =$ 2400, 180, 15 and $\dot{m}/\xi=$ $3\times10^{-3}$,
$8.5\times10^{-3}$ and $2.5\times10^{-2}$ for the three objects
respectively. Thus for $\xi$ =100 we get masses in the range $10^7 -10^9$
$M_{\odot }$ and critical accretion rates while for $\xi$=10 the masses
are higher and the accretion rates lower by a factor 10.  We conclude
that a value of $\xi$ between 10 and 100 is needed.

\section{Summary and Conclusions}

We have analized the X-ray spectra of three $\gamma $-ray loud
quasars. The main result of our analysis is that for all sources the
X-ray continuum in the 2-100 keV energy band is well represented by a
very flat ($\alpha =0.3-0.5$) power-law, without evidence for spectral
steepening at high energies.  Moreover at soft energies 0836+710 shows
evidence for either intrinsic absorption, or an extremely hard low energy
continuum. In 1510-089 a soft excess is present at $E<$ 1 keV; the presence
of a Fe line in this object is not confirmed.

By modelling the SEDs of these three sources as synchrotron and Inverse
 Compton emission from a single population of electrons with a broken
 power law energy distribution and including external seed photons for
 the IC process we estimated the physical parameters in the emission
 region and the corresponding energy transport along the jets. While the
 energy density of relativistic electrons and of the magnetic field
 are near equipartition, their energy is
 insufficient to power the observed radiation, implying that either a
 significant proton component or Poynting flux outside the emission
 region are the carriers of power.

Even the minimal jet power required by the observed radiation is very
high ($10^{45}-10^{47}$ erg s$^{-1}$) and is of the same order of that
thermally emitted in the optical-UV bands by the accretion disk.  If the
Blandford \& Znajek model for extraction of rotational energy from the
black hole is responsible for powering the jets the magnetic field at the
black hole horizon must be larger than estimated in recent works on this
highly complex and controversial issue.

\acknowledgments{We thank the $Beppo$SAX SDC for providings us with the
cleaned data. We are grateful to the anonymous referee and to Roberto
Della Ceca for helpful comments. This work was partly supported by the
Italian Ministry for University and Research (MURST) under grant
Cofin98-02-32 and by the Italian Space Agency (ASI-ARS-98-91).}

\clearpage

\vskip 1.5 truecm

\centerline{ \bf Figure Captions}

\vskip 1 truecm

\figcaption[0836_fig_nh.ps]{Fit with a Power Law and free absorption for
0836+710. \label{0836_plnhfree} }

\figcaption[multicont_gal.ps]{68, 90 and 99 \% photon index-$N_H$
confidence contours for $Beppo$SAX, ASCA and ROSAT observations of
0836+710. The horizontal solid line indicates the value of the galactic
absorption and dashed lines indicates the uncertainty range. SAX and ASCA
data clearly require extra-absorption, while ROSAT data are consistent
with a steeper, not extra-absorbed spectrum. \label{0836_multi} }

\figcaption[nh_f.ps]{ $N_H$ (as measured in the observer frame) vs. Flux
(0.1-2 keV) for different observations of 0836+710. The dashed line
indicates the Galactic absorption column. ROSAT observations are
consistent with no extra-absorption, while ASCA and $Beppo$SAX
observations require intrinsic absorption. No $N_H$-Flux correlation
seems to be present.
\label{0836_nh}}

\figcaption[figure1510.ps]{Fit of the 1510-089 spectrum with a power law
(upper panel) and with a broken power law (lower panel). The PDS/MECS
normalization is fixed to 0.85 in both fits and the PDS data are clearly
in excess to the model (see the discussion in the text).\label{1510} }

\figcaption[2230_fig.ps]{Fit with a power law and free absorption for 2230+114.
\label{2230_free} }

\figcaption[2230_cont_err.ps]{68\%, 90\% and 99\% confidence levels for $N_H$ and
spectral index for 2230+114. The solid line indicates the value of the
galactic absorption, while the dashed lines indicate the error range.
\label{2230_cont} }

\figcaption[0836_sed_new.ps]{Overall SED of 0836+710 with the spectrum calculated using the
homogeneous EC model (see text). Open circles are historical data taken
from: Kuhr et al. (1981), Wall \& Peackock (1985), Impey \& Tapia (1990),
Wiren et al. (1992), Edelson (1994) (radio), Bloom et al. (1990) (far-IR)
and Hartman et al. (1999) ($\gamma $). Triangles are simultaneous
optical data taken at the Torino Observatory. The bump in the model at
$\sim 10^{15}$ Hz is due to the black body component used to represent
the external radiation field.\label{0836_sed}}

\figcaption[1510_sed_new.ps]{Overall SED of 1510-089 with the spectrum calculated using the
homogeneous EC model (see text). Data are from: Gear et al. (1994),
Tornikoski et al. (1996), Landau et al. (1986) (radio), Pian \& Treves (1993)
(UV) and Hartman et al. (1999) (EGRET). The bump around $10^{16}-10^{17}$
Hz is due to the SSC component.\label{1510_sed} }

\figcaption[2230_sed_new.ps]{Overall SED of 2230+114 with the spectrum calculated using the
homogeneous EC model (see text). Data are taken from: Kuhr et al. (1981),
Wiren at al. (1992), Tornikoski et al. (1996) (radio), Impey \& Neugebauer
(1988) (IR), Netzer et al. (1996) (optical) and Hartman et al. (1999) (EGRET).
\label{2230_sed} }

\begin{table*}
\begin{center}
{\small
\caption{Summary of the observational characteristics of the sources
analized in this work.}
\hspace*{-1.5cm}
\begin{tabular}{lcccccccc}
\\
\hline
\hline
 & $z$ & $N_{H,gal}^a$ &$F_{\rm X}$ &$\alpha _{\rm X}$ &
$F_{\rm X} $ &$\alpha _{\rm X}$ & $F_{\gamma }$ & $\alpha _{\gamma }$\\
& & &\multicolumn{2}{c}{ROSAT$^b$} & \multicolumn{2}{c}{ASCA$^c$} &\multicolumn{2}{c}{EGRET$^d$} \\
\hline
0836+710 & 2.172 &2.78 &8.6 &0.5$\pm$0.1 & 14.0 &0.45$\pm$0.05 &10.2$\pm$1.8 &1.62 $\pm$0.16\\
 & & & 4.4 & 0.5$\pm$0.1 & & & & \\
1510-089 &0.361 & 7.88 & 6.15$\pm$0.93 & 0.9$\pm$ 0.4 & 8.6&0.30$\pm$0.06 & 18.0 $\pm$3.8 & 1.47$\pm$0.21\\
2230+114 &1.037 &5.05 &3.32$\pm$0.44 & ...  & 3.1$\pm$0.2 &0.6$\pm$0.1 & 19.2$\pm$2.8 & 1.45$\pm$0.14\\
\hline
\multicolumn{9}{l}{$a$: in units of $10^{20} cm^{-2}$. Data are from
 Dickey \& Lockman (1990) (0836+710),}\\
\multicolumn{9}{l}{
Lockman \& Savage (1995) (1510-089) and Stark et al (1992) (2230+710).} \\
\multicolumn{9}{l}{$b$: ROSAT 0.1-2 keV flux (in units of $10^{-12}$ erg
 cm$^{-2}$ s$^{-1}$) and energy index, $F_{\nu}\propto \nu ^{-\alpha}$.}\\
\multicolumn{9}{l}{From: Cappi et al. (1997)
 (0836+710, two observations), Siebert et al. (1996),}\\
\multicolumn{9}{l}{Comastri et al. (1997) (1510-089) and Brinkmann et al. (1994) (2230+114). }\\
\multicolumn{9}{l}{$c$: ASCA 2-10 keV flux (in units of $10^{-12}$ erg
 cm$^{-2}$ s$^{-1}$) and energy index.}\\ 
\multicolumn{9}{l}{From: Cappi et al. (1997)
 (0836+710), Singh et al. (1990) (1510-089) and Kubo et al. (1998) (2230+114).}\\
\multicolumn{9}{l}{$d$: EGRET flux above 100 MeV (in units of $10^{-8}$ ph
 cm$^{-2}$ s$^{-1}$) and energy index.}\\
\multicolumn{9}{l}{Average of available detections. From: Hartman et al. (1999).}\\
&&&&&&&\\
\end{tabular}
}
\end{center}
\end{table*}

{\small
\begin{table*}
\begin{center}
\caption{{\it Beppo}SAX data observation log.}
\hspace*{-1.5cm}
\begin{tabular}{lcccccccc}
\\
\hline
\hline
Date & Start& End& LECS& net cts/s$^a$& MECS &net cts/s$^b$ & PDS& net cts/s\\
 & & &Exp.(s) & &Exp.(s) & & Exp.(s)&\\
\hline
&&&&&&&&\\
\multicolumn{9}{c}{\bf 0836+710} \\ 
\hline
27-28/5/98 & 08:17:47& 08:13:14& 18209 & $0.111 \pm 0.003$ & 42640 & $0.269\pm0.003$& 16493& $0.70\pm0.05$ \\
&&&&&&&&\\
\multicolumn{9}{c}{\bf 1510-089} \\ 
\hline
3-4/8/98 & 14:23:13& 14:17:16& 15880 & $0.025\pm 0.002 $& 43870 & $0.056\pm
0.001$ & 19371& $0.31\pm0.05$\\
&&&&&&&&\\
\multicolumn{9}{c}{\bf 2230+114$^*$} \\ 
\hline
11-21/11/97 & 2:12:24& 22:51:34& 50178 & $0.033\pm 0.001 $& 103400 & $0.065\pm
0.008$ & 47834& $0.14\pm0.03$\\
\hline
\multicolumn{9}{l}{$^a$: 0.1-4 keV} \\
\multicolumn{9}{l}{$^b$: 1.8-10.5 keV, 2 MECS units} \\
\multicolumn{9}{l}{$^*$: total of 5 pointings} \\
\end{tabular}
\end{center}
\end{table*}
}

\begin{table*}
\begin{center}
\caption{Fits to {\it Beppo}SAX Data (LECS+MECS+PDS): first line single
power-law, second line broken power-law model. Errors are quoted at the
90\% confidence level for 1 parameter of interest ($\Delta \chi ^2=2.71$).}
\begin{tabular}{cccccc}
\\
\hline
\hline
$\Gamma ^a$ & \, \, $E_b^b$ & \, \, $\Gamma _h^b $\, \, & $N_H$ &
$F_{[2-10\, \rm keV]}$ & $\chi^2/$d.o.f.\\ 
      &\, \, keV & & 10$^{20}$ cm$^{-2}$ & $10^{-12}$ erg cm$^{-2}$ s$^{-1}$  &  \\
\hline
&&&&&\\
\multicolumn{6}{c}{\bf 0836+710} \\ 
\hline
$1.33\pm 0.04 $ & -& -& $8.3^{+4.7}_{-2.6}$ & 26 & 63./63\\
$0.8^{+0.4}_{-0.5}$ & $1.2\pm 0.3$& $1.31\pm 0.03$& 2.83(fixed)& 26& 63.1/62\\
\hline
&&&&&\\
\multicolumn{6}{c}{\bf 1510-089} \\ 
\hline
$1.43 \pm 0.05 $& -& -& 7.8 (fix) & 5.2 & 56.55/65\\
$2.65^{+0.63}_{-0.60} $& $1.3\pm 0.3$ & $1.39 \pm  0.08$ & 7.8 (fixed) & 5.3 & 43.1/63 \\
\hline
&&&&&\\
\multicolumn{6}{c}{\bf 2230+114} \\ 
\hline
$1.51\pm 0.04$& - &- & $7.3^{+3.8}_{-2.7}$ & 6.0 & 51.1/51 \\
$0.69^{+0.12}_{-0.08}$ & $0.8^{+0.4}_{-0.3}$ & $1.51 \pm 0.04$ & 5.04(fixed) & 6.1 & 48.0/50 \\
\hline
\multicolumn{6}{l}{$^a$: photon index, related to the spectral index by $\alpha =\Gamma-1$.} \\
\multicolumn{6}{l}{$^b$: break energy and high energy photon index for
the broken power-law model.} \\
\end{tabular}
\end{center}
\end{table*}

\begin{table*}
\begin{center}
\caption{Parameters used for the emission model described in the text.}
\begin{tabular}{cccccccc}
\\
\hline
\hline
$R$ & B & $\delta$ & $\gamma _{\rm min,inj}$ & n$_{\rm inj}$ & $L_{\rm
inj}$ & $L_{\rm BLR}$ & $R_{ \rm BLR}$\\
$10^{16}$ cm& G& & & &$10^{45}$ erg s$^{-1}$ & $10^{45}$ erg s$^{-1}$ &
$10^{18}$ cm \\ \hline
&&&&&\\
\multicolumn{8}{c}{\bf 0836+710} \\
\hline
4 & 5.9 & 18 & 50 & 3.0 & 1.48 & 3.2 & 1.8 \\
\multicolumn{8}{c}{\bf 1510-089} \\
\hline
2 & 3.1 & 18 & 60 & 3.0 & 0.01 & 0.45 & 1.0\\
\multicolumn{8}{c}{\bf 2230+114} \\
\hline
4 & 3.5 & 16 & 130 & 3.0 & 0.14 & 0.4 & 1.2 \\
\hline
\end{tabular}
\end{center}
\end{table*}

\begin{table*}
\begin{center}
\caption{Values of energy densities and jet luminosities obtained from
the emission model. See text for definitions.}
\begin{tabular}{cccc}
\\
\hline
\hline
&{\bf 0836+710} & {\bf 2230+114} & {\bf 1510-089}\\
\hline
&&&\\
$U_{\rm e}^{\prime}$ (erg cm$^{-3}$) &0.20& 0.06&0.17\\
\hline
$U_B^{\prime}$ (erg cm$^{-3}$) &1.10 &0.55 &0.38\\
\hline
$P_{\rm rad}$ ($10^{46}$ erg s$^{-1}$)& 47.9& 3.5& 0.3\\
\hline
$L_{\rm e}$ ($10^{46}$ erg s$^{-1}$)&1.00&0.23&0.18\\
\hline
$L_{B}$ ($10^{46}$ erg s$^{-1})$&4.3&1.6 &0.3\\
\hline
$L_{\rm p}^{*}$ ($10^{46}$ erg s$^{-1}$) &408&75&29.\\
\hline
$L_{\rm disk}$ ($10^{46}$ erg s$^{-1}$)  &10.0 &2.00 &0.50\\
\hline
\multicolumn{4}{l}{$^*$: calculated for $n_p=n_e$} \\
\end{tabular}
\end{center}
\end{table*}

\clearpage

\begin{figure}
\centerline{\plotone{f1}}
\end{figure}

\clearpage

\begin{figure}
\centerline{\plotone{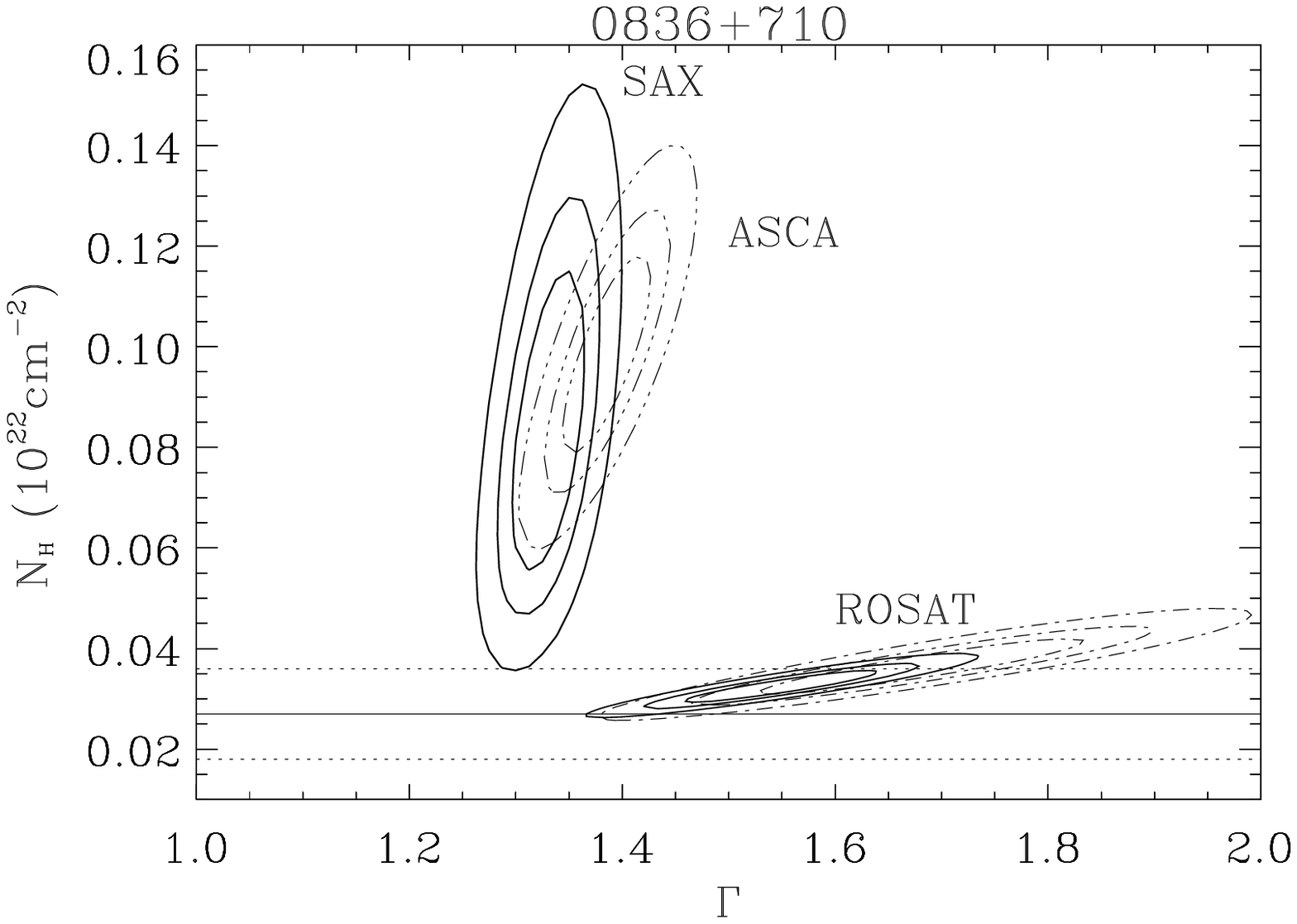}}
\end{figure}

\clearpage

\begin{figure}
\centerline{\plotone{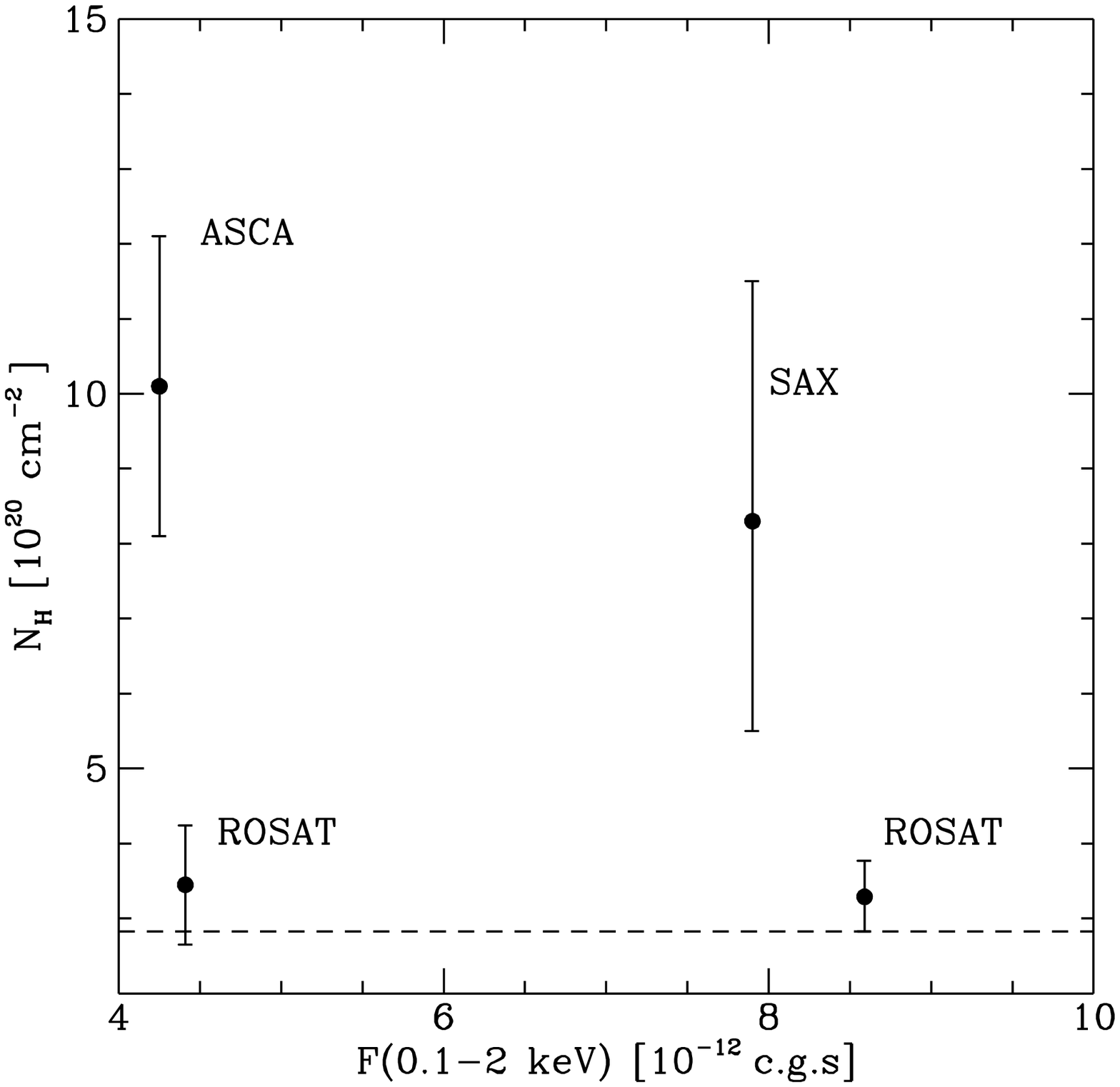}}
\end{figure}

\clearpage

\begin{figure}
\centerline{\plotone{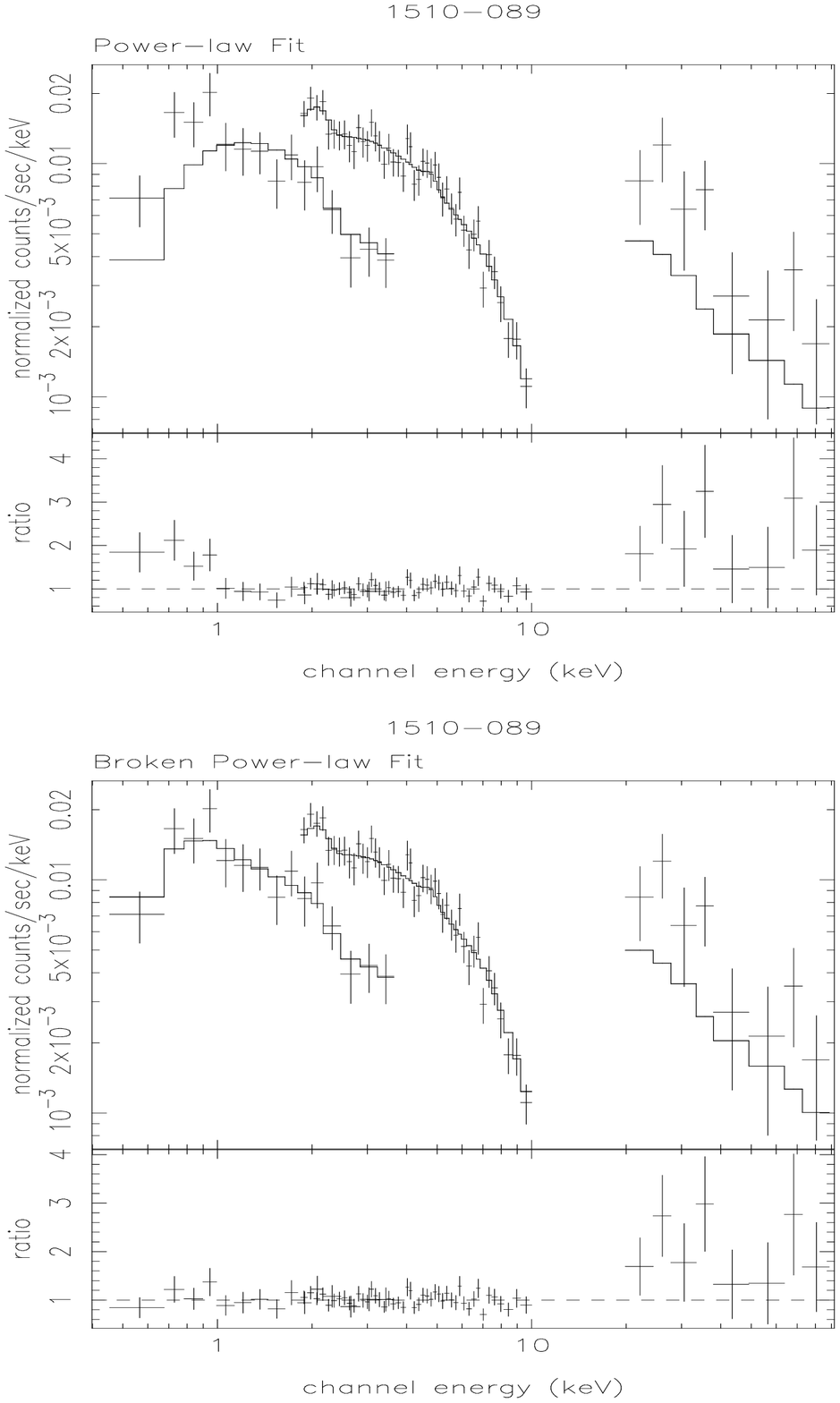}}
\end{figure}

\clearpage

\begin{figure}
\centerline{\plotone{f5}}
\end{figure}
\clearpage

\begin{figure}
\centerline{\plotone{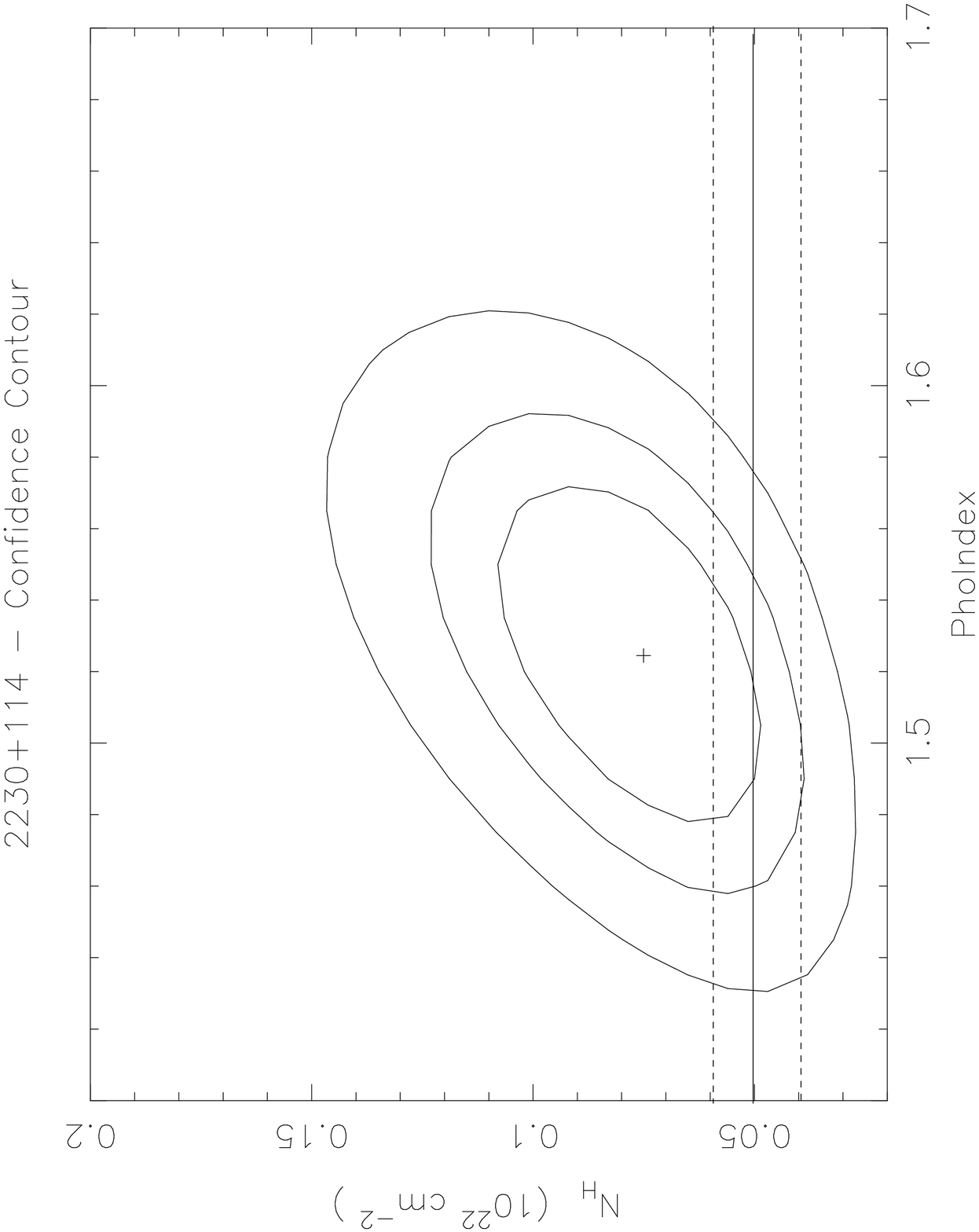}}
\end{figure}
\clearpage

\begin{figure}
\centerline{\plotone{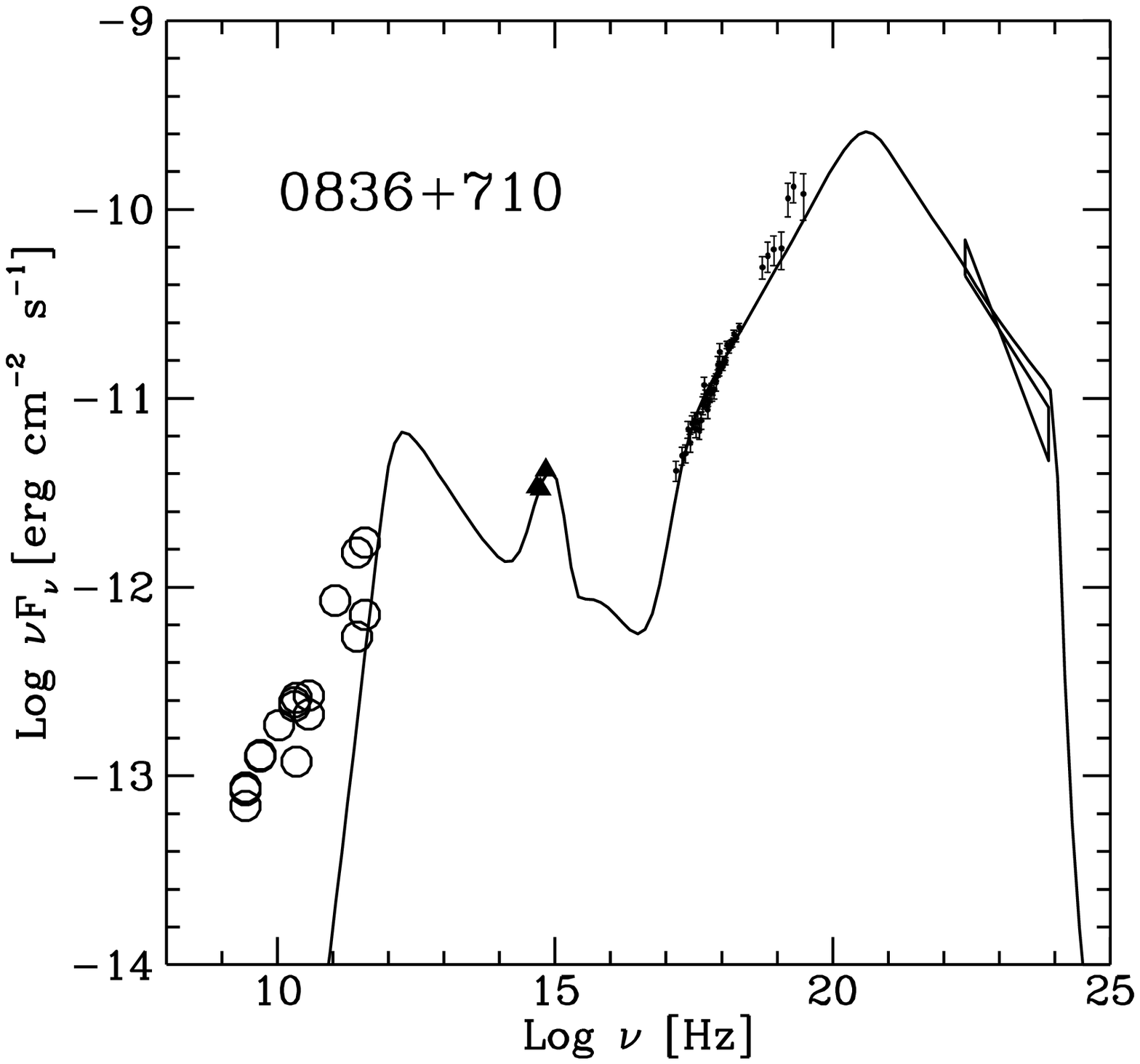}}
\end{figure}
\clearpage

\begin{figure}
\centerline{\plotone{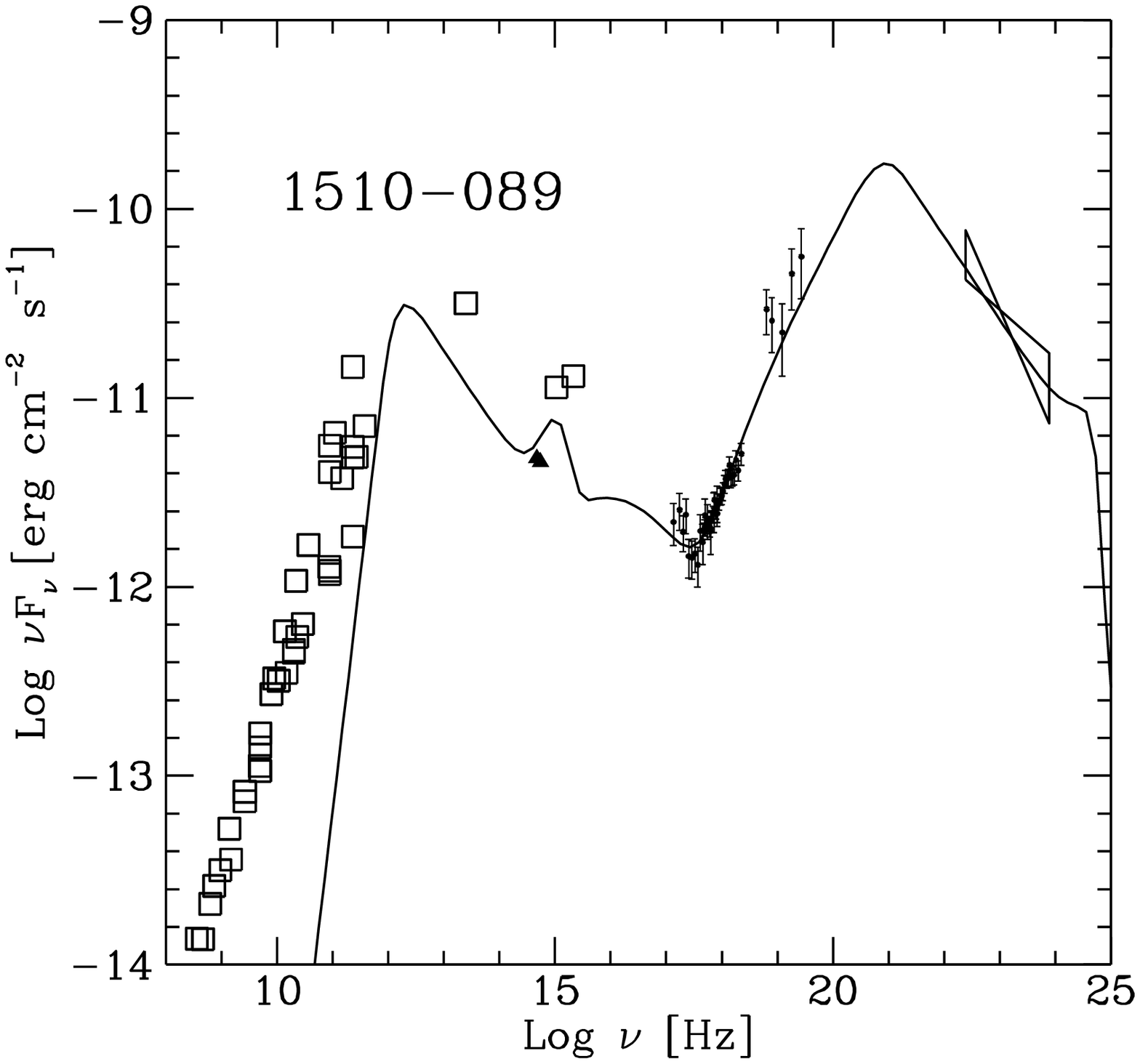}}
\end{figure}
\clearpage

\begin{figure}
\centerline{\plotone{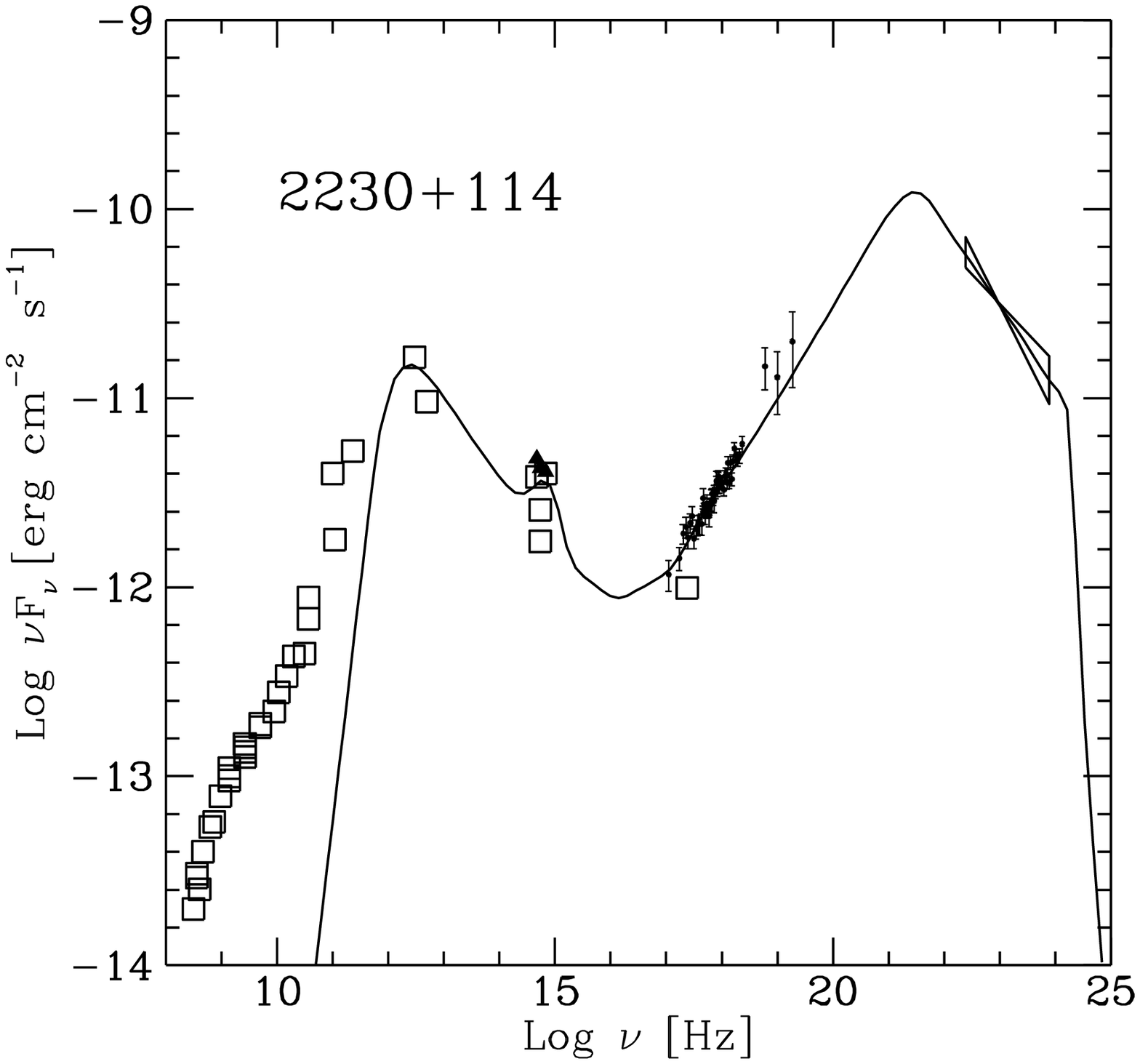}}
\end{figure}

\end{document}